\documentclass[conference]{IEEEtran}
\IEEEoverridecommandlockouts
\usepackage{cite}
\usepackage{amsmath,amssymb,amsfonts}
\usepackage{algorithmic}
\usepackage{graphicx}
\usepackage{textcomp}
\usepackage{xcolor}
\usepackage{soul}
\usepackage[algo2e]{algorithm2e}
\usepackage{algorithm}

\usepackage{braket}
\usepackage{amsmath,amsbsy,amsfonts,amssymb,amsthm,commath}

\def\BibTeX{{\rm B\kern-.05em{\sc i\kern-.025em b}\kern-.08em
    T\kern-.1667em\lower.7ex\hbox{E}\kern-.125emX}}
\begin{document}

\title{Large-scale Quantum Approximate Optimization via Divide-and-Conquer}

\author{\IEEEauthorblockN{Junde Li}
\IEEEauthorblockA{
\textit{Pennsylvania State University}\\
jul1512@psu.edu}
\and
\IEEEauthorblockN{Mahabubul Alam}
\IEEEauthorblockA{\textit{Pennsylvania State University} \\
mxa890@psu.edu}
\and
\IEEEauthorblockN{Swaroop Ghosh}
\IEEEauthorblockA{
\textit{Pennsylvania State University}\\
szg212@psu.edu}

}

\IEEEaftertitletext{\vspace{-1.5\baselineskip}}
\maketitle

\begin{abstract}
Quantum Approximate Optimization Algorithm (QAOA) is a promising hybrid quantum-classical algorithm for solving combinatorial optimization problems. However, it cannot overcome qubit limitation for large-scale problems. Furthermore, the execution time of QAOA scales exponentially with the problem size. We propose a Divide-and-Conquer QAOA (DC-QAOA) to address the above challenges for graph maximum cut (MaxCut) problem. The algorithm works by recursively partitioning a larger graph into smaller ones whose MaxCut solutions are obtained with small-size NISQ computers. The overall solution is retrieved from the sub-solutions by applying the combination policy of quantum state reconstruction. 
Multiple partitioning and reconstruction methods are proposed/ compared. 
DC-QAOA achieves 97.14\% approximation ratio (20.32\% higher than classical counterpart), and 94.79\% expectation value (15.80\% higher than quantum annealing). DC-QAOA 
also reduces the time complexity of conventional QAOA from exponential to quadratic.

\end{abstract}


\section{Introduction}

\begin{figure*}
\centering
\includegraphics[width=17.5cm]{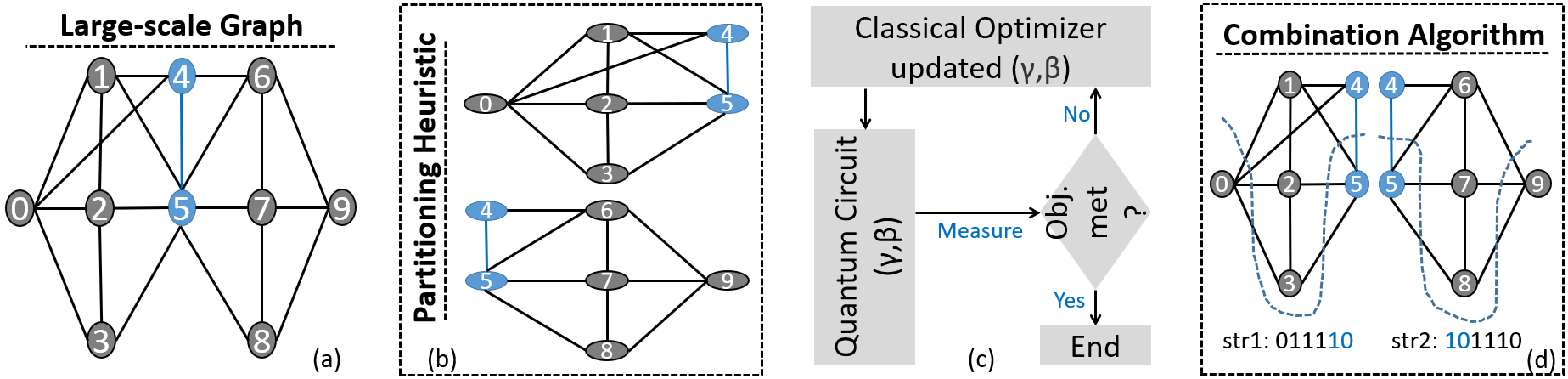}
\caption{Finding MaxCut for a large-scale graph using the proposed DC-QAOA algorithm: (a) a graph with 10 nodes that is unsolvable on a small quantum computer (say with 6 qubits); (b) graph partitioning heuristic (proposed \textit{LGP} policy) splits the larger graph into two subgraphs with almost equal number of nodes (say only nodes 4 and 5 are shared in this graph); (c) MaxCut of each subgraph is solved by the hybrid QAOA, where each cost Hamiltonian is evolved and measured on the small quantum hardware and variational gate parameters are classically updated during every iteration; (d) the solution for original large graph is retrieved using the combination algorithm (proposed \textit{QSR} policy) which takes the sampling distribution of each subgraph solution as input.}
\label{flow}
\vspace{-5mm}
\end{figure*}

Quantum Computing (QC) 
can offer unique advantages over classical computing 
in many important areas such as, machine learning and optimization \cite{peruzzo, biamonte, hu}. 
Approximate hybrid quantum-classical algorithms such as, Quantum Approximate Optimization Algorithm (QAOA) \cite{qaoa} is proposed to work in presence of noise sources in near-term NISQ machines. QAOA contains a variational quantum circuit and a classical optimizer (Figure \ref{flow}(c)) to approximately solve combinatorial optimization (CO) problems like MaxCut \cite{qaoa}. However, QAOA is claimed to be 20X slower than its classical counterpart for finding MaxCut of a 20 node graph since the number of operations and circuit depth $p$ in quantum simulation both scale exponentially with graph size \cite{guerreschi}. We propose a divide-and-conquer QAOA (DC-QAOA) to solve large-scale combinatorial problems and to reduce the QAOA computational complexity. All experiments on quantum optimization are conducted on the representative MaxCut problems. 


\textbf{Proposed idea:} DC-QAOA (Figure \ref{flow}) has two components:

\textit{(i) Graph splitting} which recursively divides the large-scale graph with nodes greater than the qubit size of a target quantum hardware into subgraphs by using \textit{Large Graph Partitioning (LGP)} policy. The MaxCut of subgraphs can be solved efficiently using QAQA on quantum simulator or real quantum computers. Solution of the original graph is retrieved using the combination policy of \textit{Quantum State Reconstruction (QSR)}.
If a pair of subgraphs have a single common node and no crossing edges, that would be ideal scenario. Each subgraph MaxCut can be solved individually and sub-solutions can be combined to determine the MaxCut of the large graph. However, real cases can share more nodes that incur extra qubit overhead and also reduce the accuracy of quantum state reconstruction. The objective of \textit{Large Graph Partitioning} policy is to minimize the crossing edges and number of common nodes. 
Figure \ref{flow}(a) shows a large-scale 10-node graph to be solved on a 6-qubit NISQ machine. Our \textit{Naive LPG} approach iteratively increases the number of common nodes while avoiding crossing edges until a nodal-separation path is found (refer to Algorithm \ref{lgp}). With \textit{Naive LPG}, path ($4$, $5$) is one of the valid separations for graph in Figure \ref{flow}(a), where nodes $4$ and $5$ are shared. Partitioning with minimized number of common nodes and crossing edges takes up least qubit resources by our DC-QAOA algorithm. We propose the \textit{Node Redundancy Level (NRL)} metric to evaluate the effectiveness of the splitting graph. $NRL$ is the ratio of the sum of subgraph node size to the original graph node size. 

\textit{(ii) Quantum state reconstruction} that aims to reconstruct the full quantum state from experimentally accessible measurements for a quantum system \cite{torlai, leonhardt}. Multiple reconstruction schemes existing in literature, such as, Born's rule in probability representation and Wigner function \cite{leonhardt}. However, the \textit{Quantum State Reconstruction (QSR)} policy used in the current study is different from the conventional reconstruction concept since we intend to reconstruct a larger quantum state from available measurements of its quantum sub-systems. \textbf{One may intuitively doubt the effectiveness of \textit{QSR} policy for combining subgraph solutions because entanglement entropy cannot be reconstructed from two separate quantum systems. Fortunately, qubit entanglement created by \textit{Controlled-NOT (CNOT)} gates between subgraphs is bypassed by our \textit{LGP} policy which avoids crossing edges}.
The combination policy only applies to quantum algorithms like QAOA due to the requirement of state measurements.

After the optimal gate parameters are found in Figure \ref{flow}(c), we obtain the quantum state frequency distribution by taking certain number of measurements from each subgraph solution space. The larger quantum state is reconstructed by strictly complying with the \textit{combination criterion} of sharing the same cut set $S$ for each common node. For instance, $str1$ and $str2$ in Figure \ref{flow}(d) are two highly sampled solution strings from their subgraphs, respectively, and node $4$ belongs to the same cut set $S$ and node $5$ belongs to the complementary set $\bar{S}$ from each sub-solution. Therefore, the combined solution string $0111101110$ is the optimal cut solution for original graph in Figure \ref{flow}(a). Once the \textit{combination criterion} is satisfied, multiple reconstruction schemes can be adopted, such as, by taking the minimum of the two solution string frequencies, or taking their product, etc. We evaluate the reconstruction quality of each scheme using the expectation value and \textit{Kullback–Leibler divergence} as metric relative to the estimated quantum state from the original graph.

\textbf{Contributions:} We propose, (a) a novel DC-QAOA algorithm for solving large-scale MaxCut problems with exponential speedup
suitable for small NISQ machines; (b) novel \textit{Quantum State Reconstruction} policy for approximately reconstructing the quantum state of a large system from its sub-systems; (c) evaluation of \textit{Large Graph Partitioning} scheme using our \textit{Node Redundancy Level} metric. 
Our algorithm only applies to input graphs with connectivity less than maximum allowed qubit size. However, qubits provided by near-term quantum computers or simulator are reasonably sufficient.

\section{Related Work and Motivation}

\textbf{Solving large problems on small NISQ hardware:} Dunjko et al. \cite{dunjko} proposed a hybrid algorithm 
for solving 3SAT problems on quantum computers with limited number of qubits. Ge $\&$ Dunjko \cite{ge2020hybrid} proposed another hybrid algorithm to enhance Eppstein's algorithm for finding cubic Hamiltonian circle in degree-3 graphs. However, these hybrid algorithms do not apply to 
the MaxCut problem addressed in this paper.

\begin{figure}
\centering
\includegraphics[width=7.5cm]{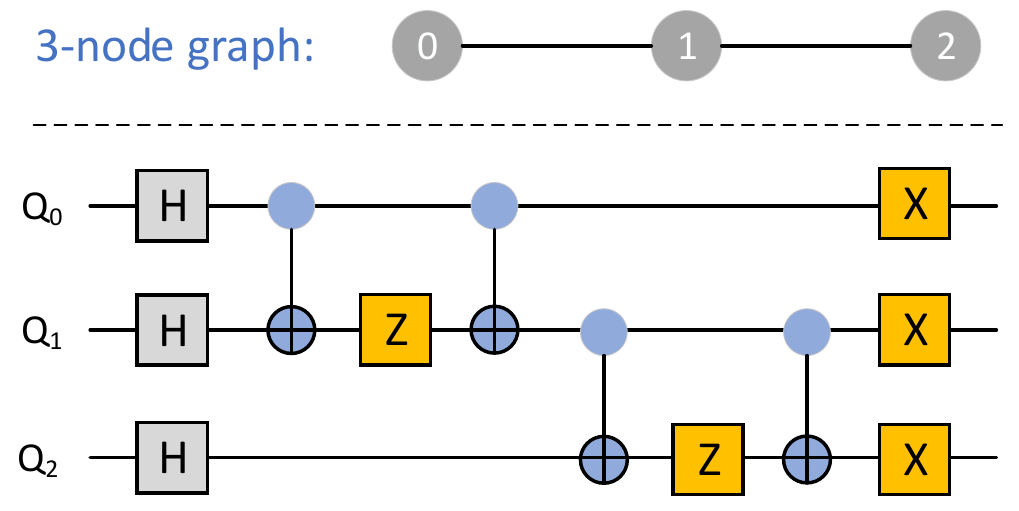}
\caption{QAOA circuit after graph is mapped on a small quantum system for getting MaxCut solution. Gate parameters $(\gamma_i, \beta_i)$ are not shown here.}
\label{qc_basics}
\vspace{-5mm}
\end{figure}

\textbf{QAOA:} 
The QAOA circuit (see Figure \ref{qc_basics}) can be configured with $p$ layers by repeating the non-H gates for $p$ times, while the gate parameter pair $(\gamma_i, \beta_i)$ corresponds to layer $i$. The final quantum state is measured and the expectation value is calculated accordingly. Classical optimizer updates the gate parameters until convergence is reached (Figure \ref{flow}(c)).

\begin{figure}
\centering
\includegraphics[width=8cm]{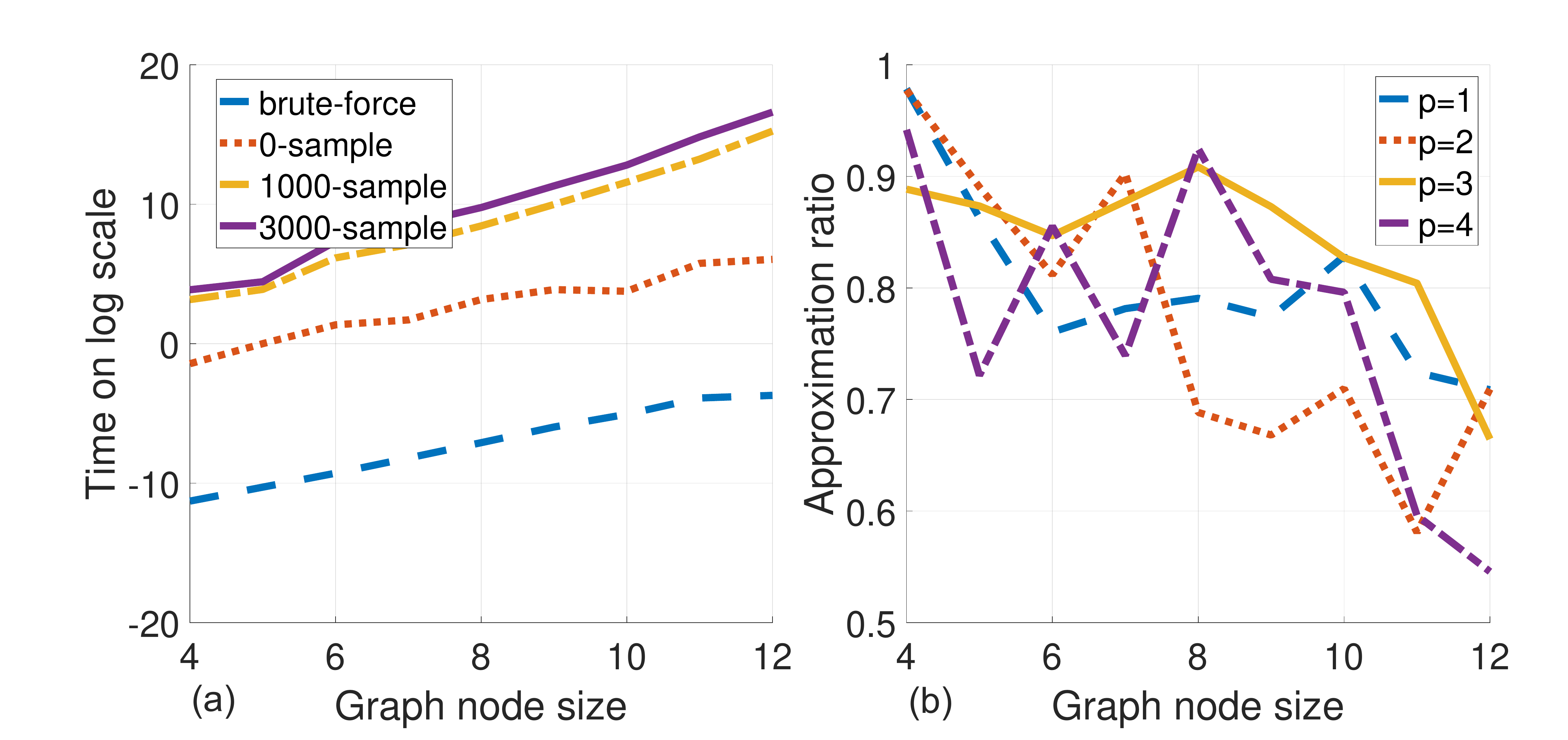}
\caption{(a) Time on log scale (base 2) elapsed for solving MaxCut of the graphs with node size from 4 to 12 by classical brute force and QAOA with three different sampling configurations; (b) approximation ratio of QAOA with 0-sample configuration at different quantum circuit depth from $p=1$ to 4.}
\label{qaoa}
\vspace{-5mm}
\end{figure}

The classical brute-force method runs significantly faster than the hybrid QAOA algorithm configured with 10 iterations without sampling of quantum states (Figure \ref{qaoa}(a)). The QAOA algorithm with 3000 samples takes 27.77 hours for solving the MaxCut of a 12-node graph, while QAOA without sampling takes 51.79s compared to 0.08s by brute force method. All the four solver schemes scale exponentially with graph size. Figure \ref{qaoa} (b) shows the decreasing approximation ratios of QAOA in general for all 4 different layers, relative to the ground truth cut solutions from brute force. 

 \textbf{Motivation:} Currently, QAOA is far from achieving quantum advantage for MaxCut problems compared to classical counterparts. We aim to speed up QAOA algorithm exponentially for large-scale graphs with up to 100 nodes.
 Let us assume a large-scale graph has $N$ nodes and maximum available quantum qubit size is $K$. QAOA running on a quantum simulator has $\mathcal{O}(2^N)$ time complexity, while DC-QAOA achieves $\mathcal{O}(N^{2K} (1+1/K) N)$ complexity, where each small problem takes constant time.


\section{Proposed Methodologies}

\subsection{Large Graph Partitioning}
\textbf{Background:} 
Graph partitioning can be classified into \textit{edge separator} and \textit{node separator}. 
Given a graph $G(N, E)$, \textit{node separator} $N_s$ separates the graph $G$ into two disconnected components $G_1$ and $G_2$ if all nodes in $N_s$ and incident edges are removed. 
We note that \textit{node connectivity} algorithm is most related to our partitioning policy. \textit{Node connectivity} is the minimum number of nodes that must be removed to disconnect an input graph. However, \textit{node connectivity} algorithm cannot be directly applied to our partition task since the found separation nodes are not shared by two subgraphs.

\textbf{LGP description:} 
Let $G(V, E)$ be an undirected graph with $n$ vertices an $m$ edges. All the edges are unweighted or considered with a weight of 1. $LGP$ partitions $G$ into exactly two subsets, $S$ and $\bar{S}$ no larger than a given maximum size of $k$ (i.e., available qubit size in NISQ computers). 
Note that at least one shared node is required such that no edges are missing after each partition. The path connecting the shared nodes $(s_1, s_2, ..., s_i)$ forms the final separation line. The metric \textit{Node Redundancy Level (NRL)} is defined as the ratio of sum of node sizes from subgraphs to $n$ for measuring the quality of partition. We propose the \textit{Naive LGP} scheme for approximately minimizing the cut cost in Algorithm \ref{lgp}.

\begin{algorithm}[H]
\SetAlgoLined
\caption{Naive LGP Heuristic (NLGP)}\label{lgp}
\textbf{Input:} $G(V, E)$ - connected graph;
$k$ - maximum node size.\\
\textbf{Output:} $S = \{S_1, S_2\}$ - final set of subgraphs.\\
counter = 1 \\
\While{counter $<$ k}{
    paths = [] \\
    \uIf{counter = 1}{
        paths $\gets$ V
    }
    \Else{
        \textit{\# try (counter-1) nested edges loops for appending\\          \# possible paths (loops not being displayed here)} \\
        pos\_path $\gets$ [$v_1$, $v_2$, ..., $v_{counter}$] \\
        \If {len(unique(pos\_path)) = counter}{
            paths.append(pos\_path)
        }
    }
    
    \textit{\# check suitable separation path} \\
    \For {p in paths}{
        G$_1$ $\gets$ G(V-p, E) \textit{\# remove p and incident edges} \\
        S $\gets$ dfs\_connected\_components(G$_1$)\\
        \If{len(S) = 2}{
            \Return{S} \textit{\# exactly 2 subsets $S_1$ and $S_2$}
        }
    }
    counter ++
}
print (G has connectivity above $k$)\\
\Return\{\}
\end{algorithm}

\textbf{Proposed separation paths:} \textit{NLGP} heuristic proposes all possible separation paths with $counter$ number of nodes. For $counter=1$, the $paths$ list incorporates each node. Each edge $(u, v)$ is proposed for finding path candidates with two nodes for $counter=2$. For $2<counter<k$, code block in else statement simplifies the $counter-1$ nested edges loops that is the time intensive part $\mathcal{O}(m^k)$ in worst case of \textit{NLGP}.

\textbf{Select workable path:} Each path candidate is a set $p$ composed of a sequence of nodes. A new graph $G_1$ is reduced from $G$ by removing all nodes in $p$ and incident edges. The new graph is fed into Depth-first Search (DFS) algorithm for finding all connected components. The right separation path is found until DFS outputs exactly 2 connected components.

\begin{figure*}
\centering
\includegraphics[width=16.5cm]{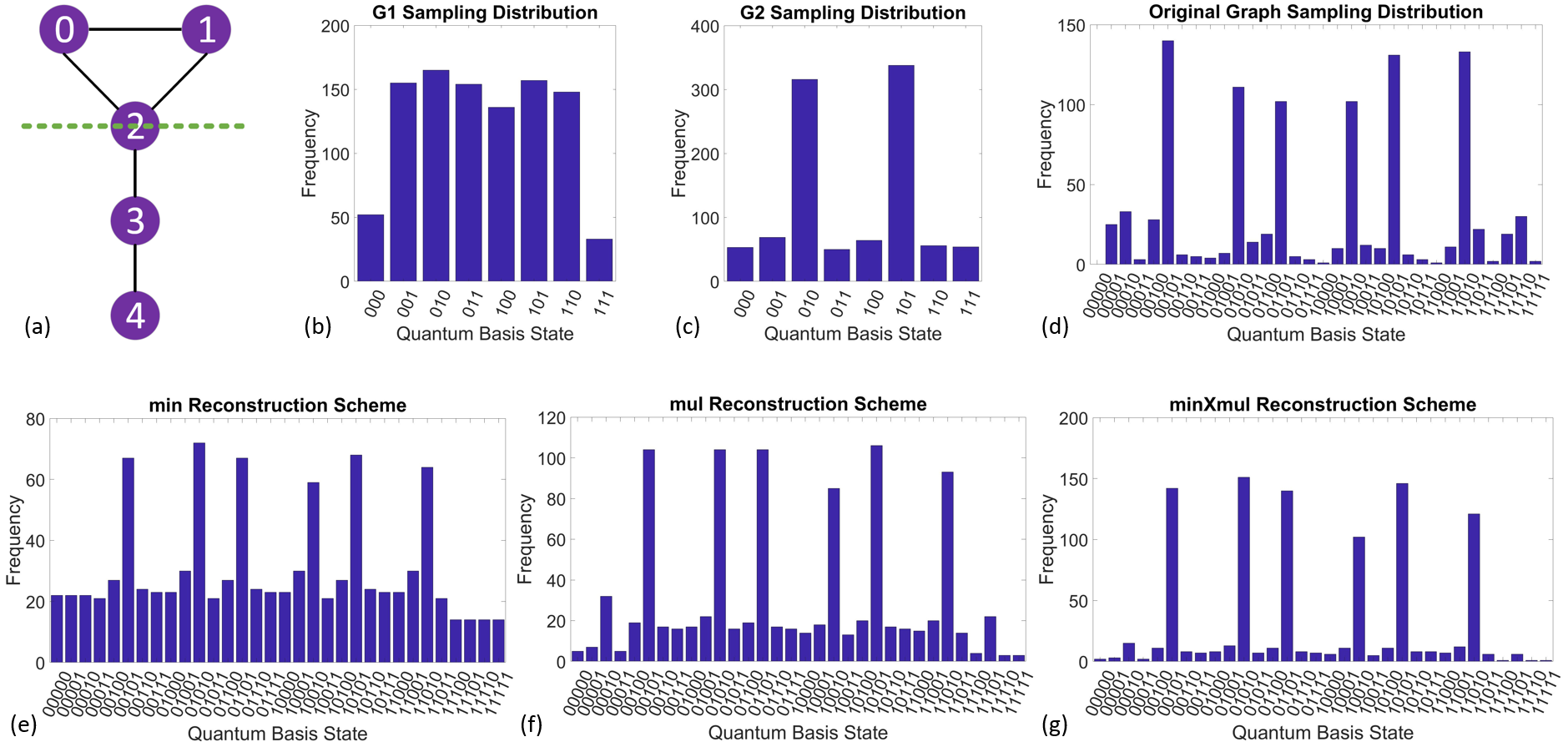}
\caption{Quantum state reconstruction demonstration with 1000 samples and circuit depth of 3: (a) a toy graph $G$ partitioned into $G_1$ and $G_2$ by a single common node 2; (b-c) quantum state distributions of subgraphs $G_1$ and $G_2$ on small quantum hardware (6 optimal solution strings for $G_1$ and 2 optimal ones for $G_2$); (d) quantum state distribution of original graph on a larger hardware (6 optimal solution strings); (e-g) reconstructed quantum state distributions of three \textit{QSR} schemes of \textit{min}, \textit{mul} and \textit{minXmul}, respectively (sampling frequency of each scheme is re-scaled with 1000 total samples).}
\label{qsr_schemes}
\vspace{-5mm}
\end{figure*}

The \textit{NLGP} heuristic only separates input graph $G$ into two subsets $S_1$ and $S_2$, thus subgraphs containing node set $S_1$ or $S_2$ will execute the same \textit{NLGP} heuristic until all final subsets are within maximum node size $k$. Two subsets after every partition may not have equal node size, however, the Naive heuristic still does not affect the partition efficiency if the recursive process is not employed in parallel. Like usual \textit{node connectivity} algorithm, \textit{NLGP} cannot separate input graph $G$ with connectivity greater than $k$. 

\subsection{Quantum State Reconstruction}

\textbf{Background:} 
Unlike conventional quantum state tomography \cite{yang, torlai, Weigert2009}, our \textit{QSR} policy approximately reconstructs the quantum state of a larger system from state distributions of its two quantum sub-systems. The set of measurement operators must span all possible Hermitian measurements to accurately reconstruct the quantum state. However, the number of measurements grow exponentially with the number of qubits preventing the quadratic-time execution for large-scale graphs. We only sample the measurements for a certain number of times to circumvent the exponential growth rate.

\textbf{QSR description:} \textit{LGP} policy avoids crossing edges between two subgraphs, which makes the state construction of a parent system possible by getting rid of qubit entanglement between sub-systems. The reconstruction quality is quantitatively measured by the similarity between reconstructed quantum state distribution and original quantum state using KL divergence since our objective is to reconstruct quantum state as close as to parent quantum state.

Algorithm \ref{qsr} shows the logic for combining bit string solutions from the subgraphs. 
Each bit string solution map keeps the pairs of quantum state and its corresponding sampling frequency. The validity code line checks if the aforementioned \textit{combination criterion} is met before the original bit string is reconstructed. Multiple \textit{QSR} schemes may apply here such as, \textit{min}, \textit{mul}, \textit{sum}, etc. Algorithm \ref{qsr} takes the example of \textit{min} scheme where the minimum frequency counts is applied. \textit{Sum} returns the sum of counts from two solution maps, while \textit{mul} returns the multiplication. It is worth mentioning that not all bit string solutions are sampled most frequently all the time because of the probabilistic nature of quantum measurement. To counteract this unwanted effect, an important re-ranking mechanism is followed by re-ranking the frequency counts based on the calculated cut sizes of their corresponding bit strings. The complexity of re-ranking mechanism (not shown in Algorithm \ref{qsr}) is $\mathcal{O}(2^k)$ in worst case.

To compare the reconstruction quality of several possible \textit{QSR} schemes, we conduct several experiments on a toy 5-node graph, shown in Figure \ref{qsr_schemes} (a), where the single node $2$ forms the separation path. The quantum basis state of subgraph $G_1$ (with node 0, 1 and 2) and $G_2$ (with rest of the nodes) is sampled 1000 times from quantum system with circuit depth of 3, respectively (Figure \ref{qsr_schemes}(b-c)). Figure \ref{qsr_schemes}(d) displays the basis state sampling distribution for the original graph. After extensive search of possible schemes, \textit{min}, \textit{mul} and \textit{minXmul} (the product of \textit{min} and \textit{mul}) are found to be good candidates (Figure \ref{qsr_schemes}(e-g)). Ideally, all measurement results fall on the optimal 6 basis states corresponding to 6 optimal MaxCut solutions for the toy graph with equal likelihood of $16.67\%$ (in absence of quantum noise and infinite number of samples). Intuitively, \textit{min} scheme prevents unbalanced distribution for further reconstruction compared to the original distribution however, non-optimal states are highly weighted. Non-optimal states are relatively well kept by \textit{mul} scheme, but frequencies of optimal states are not as good as \textit{minXmul} scheme.

The reconstruction quality is quantitatively measured with \textit{KL divergence} and the expectation value metrics. \textit{Mul} scheme achieves the lowest entropy ($0.164$) followed by \textit{minXmul} ($0.182$), and \textit{min} achieves the highest divergence ($0.332$), relative to the original distribution. Due to the uncertain nature of parent state distribution, we consider \textit{mul} and \textit{minXmul} with the same level of reconstruction quality. Note that, lower entropy score only indicates closer reconstruction to parent state distribution, but not better quantum expectation value in general. Quantum expectation value is defined as the average cut size of certain input graph, and also approximation ratio if divided by the known MaxCut. Therefore, expectation value of parent quantum system and all \textit{QSR} schemes are further evaluated. Expectation value calculated through the sampling frequencies reaches 3.52 (88\% approximation ratio) for original quantum system, with 4 being MaxCut for graph in Figure \ref{qsr_schemes}(a). Apparently \textit{min} scheme achieves the lowest value of 2.97 (74.27\% approximation ratio), while \textit{minXmul} scheme achieves the highest value of 3.72 (93\% approximation ratio). By trying more complicated \textit{QSR} schemes we can achieve even higher approximation ratio. 

\begin{algorithm}[H]
\SetAlgoLined
\caption{Quantum State Reconstruction (QSR)}\label{qsr}
\textbf{Input:} $g_1$, $g_2$ - two subgraphs with few common nodes;\\
$str\_cnt1$, $str\_cnt2$ - string solution maps for $g_1$ and $g_2$.\\
\textbf{Output:} $com\_cnt$ - combined bit string solution map.\\\
com\_cnt $\gets$ \{\}\\
common\_node $\gets$ intersection($g_1.nodes$, $g_2.nodes$) \\
\For {(str1, cnt1) in str\_cnt1}{
    \For {(str2, cnt2) in str\_cnt2}{
        \textit{\# check equality for each bit in common nodes} \\
        validity $\gets$ [str1[v] = str2[v] for v in common\_node]\\
        \If {False not in validity}{
            com\_str $\gets$ $''$ \textit{\# initialized with empty string}\\
            \For{i in unique($g_1.nodes$ + $g_2.nodes$)}{
                \uIf{i in $g_1.nodes$}{
                    com\_str.join(str1[i])
                }
                \Else{
                    com\_str.join(str2[i])
                }
            }
            \textit{\# min reconstruction scheme here} \\
            com\_cnt[com\_str] $\gets$ min(cnt1, cnt2)
        }
    }
}
\textit{\# sort string-count map by counts in reverse order} \\
com\_cnt $\gets$ sorted(com\_cnt) \\
\Return com\_cnt
\end{algorithm}

\subsection{DC-QAOA}
\label{sec:dc-qaoa}

DC-QAOA (Algorithm \ref{dc-qaoa}) 
recursively divides large-scale graph with \textit{LGP} policy and combines sub-solutions with \textit{QSR} policy via divide-and-conquer paradigm. 
The input graph is partitioned into exactly 2 subgraphs with \textit{LGP} policy if its node size is greater than allowed qubit size $k$, or directly solved by conventional QAOA otherwise. Each subgraph recursively call the DC-QAOA function until all their sub-solutions $str\_cnt$ are returned. Note that, two sub-solutions employ a weighting mechanism before being combined which assigns weights based on their respective node sizes to each frequency count. The weighting mechanism provides a means to allocating higher priority to the subgraph with more nodes, though both are within $k$-node. A parent solution (quantum state) $out\_cnt$ is further reconstructed by calling the \textit{LGP} policy with subgraphs and their sub-solutions as arguments. For output $out\_cnt$ directly returned from QAOA, solution pairs are sorted by frequency count, using the same sorting scheme employed by Algorithm \ref{qsr}.

\begin{figure*}
\centering
\includegraphics[width=17.5cm]{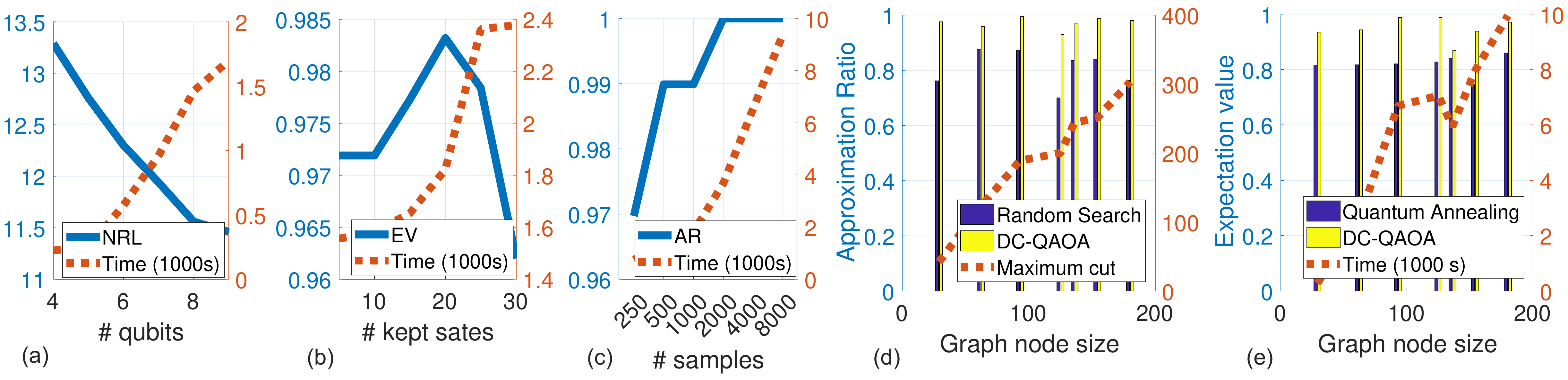}
\caption{(a) Smaller allowed qubit size increases node redundancy level (NRL) but reduces overall run time; (b) number of retained quantum basis states affects the expectation value (EV); (c) larger number of measurements (samples) increases approximation ratio (AR); (d) approximation ratio comparison with representative random search classical algorithm, together with ground truth MaxCut from Gurobi optimizer; (e) expectation value comparison with quantum annealing (from D-Wave Systems) together with run time of DC-QAOA for 7 graph instances.}
\label{res}
\vspace{-5mm}
\end{figure*}

\begin{algorithm}[H]
\SetAlgoLined
\caption{Divide-and-Conquer QAOA (DC-QAOA)}\label{dc-qaoa}
\textbf{Input:} $g$ - input graph; $p$ - quantum circuit depth; $t$ - top $t$ samples; $s$ - number of samples; $k$ - max qubit size. \\
\textbf{Output:} $out\_cnt$ - bit string solution map for $g$.\\
\uIf{len(g.nodes) $>$ k}{
    \textit{\# get exactly two subgraphs with \textit{LGP} policy} \\
    $g_1$, $g_2$ $\gets$ $LGP(g, k)$ \\
    common\_node $\gets$ intersection($g_1.nodes$, $g_2.nodes$) \\
    $str\_cnt1$ $\gets$ DC-QAOA$(g_1)$ \\
    $str\_cnt2$ $\gets$ DC-QAOA$(g_2)$ \\
    \textit{\# weighted string-count maps by node size} \\
    $str\_cnt1$ $\gets$ \{k: v*len(k) for k, v in $str\_cnt1$\} \\
    $str\_cnt2$ $\gets$ \{k: v*len(k) for k, v in $str\_cnt2$\} \\
    \textit{\# reconstruct string-count map with \textit{LGP} policy} \\
    $out\_cnt$ $\gets$ $QSR(g_1, g_2, str\_cnt1, str\_cnt2)$
}
\Else{
    $out\_cnt$ $\gets$ qaoa\_maxcut$(g, p)$ \\
    \textit{\# sort string-count map by counts in reverse order} \\
    $out\_cnt$ $\gets$ sorted$(out\_cnt)$
}
\textit{\# retain only top $t$ (str, cnt) pairs by sorted order} \\
$out\_cnt$ $\gets$ abridged$(out\_cnt, $t$)$ \\
\textit{\# rescale total number of counts to $s$ or around} \\
cnt\_sum $\gets$ the sum of all counts in $out\_cnt$ \\
$out\_cnt$ $\gets$ \{k: int(s*v/cnt\_sum) for (k, v) in $out\_cnt$\}

\Return $out\_cnt$
\end{algorithm}

At the end, we employ selection and re-scaling. The selection mechanism only retains the top $t$ sorted str-cnt solution pairs. If some str-cnt pairs with zero count are within the top $t$, only pairs with non-zero count are retained. The selection mechanism only keeps top cut solutions which prevents exponential scaling with the number of quantum logical states. It also reduces the noise caused by non-optimal solutions for further parent state reconstruction. The re-scaling step is applied for capping the sum of counts at $s$ otherwise the reconstructed state distribution may become extremely unbalanced (especially for \textit{mul QSR} scheme) for further reconstruction.

\subsection{DC-QAOA Complexity}
The worst case time complexity of \textit{Naive LGP} policy is $\mathcal{O}(m^k)$. However, graph partitioning solution is usually found with only few iterations indicating a way more efficient average-case run time. Exponential speedup is achieved primarily due to the selection mechanism in \textit{QSR} policy which guarantees constant time for each parent state reconstruction. Each divided small problem is solved by conventional QAOA only in constant time. Assuming \textit{LGP} policy splits graph into subgraphs with almost equal size, the overall DC-QAOA run time is dominated by partitioning $T(\textit{LGP})$. Each partition shares at most $k$ common nodes and there are approximately $N/k$ partitions in total. Thus the overall approximate run time is $\mathcal{O}(N^{2K} (1+1/K) N)$.

\section{Evaluation}

\subsection{Experimental Setup}
A set of large-scale graphs with varied size is randomly generated using NetworkX package for evaluating DC-QAOA performance.
We conduct a series of sensitivity analyses for choosing appropriate parameters of $t$, $s$, and $k$ based on a single 100-node graph. The baseline algorithm chooses $t=20$, $s=1000$ and $k=8$. The tuning results are shown in Figure \ref{res}. 
From Figure \ref{res}(a), larger number of allowed qubits significantly decreases node redundancy level by separating the input graph into less number of sub-problems, however it increases the overall run time due to exponential growth rate of QAOA. The algorithm is configured with $k=7$ to solve MaxCut of input graph with larger connectivity. As more number of quantum basis states are retained, the likelihood of these solution strings being optimal reduces. If the number of kept states are too small it is subjected to quantum state reconstruction noise. We pick $t=20$ where the expectation value is maximized and the run time is relatively low (Figure \ref{res}(b)). The approximation ratio in Figure \ref{res}(c) depends on the number of samples in general because quantum state reconstructed from a larger number of measurements reflects the real state more accurately. Only 2000 samples are taken due to the exponentially growing sampling time. During all these 18 experiments, the least approximation ratio of 96.97\% is obtained only once at $s=250$, while $98.99\%$ is achieved for 7 times. For all other cases, we noted $100\%$ approximation ratio. Therefore, we assume that DC-QAOA is able to generate (near-)optimal MaxCut solution for large-scale graphs as long as enough samples (say, 5000) are measured. 

\subsection{Comparison}
For fair evaluation of DC-QAOA, we also implement several counterpart classical MaxCut solvers 
and quantum annealing.

\textbf{Classical counterparts:} MaxCut problem has the reduced form of QUBO framework. Therefore, classical QUBO solvers such as, Random Search \cite{friedrich2019greedy} and Gubori solver \cite{gurobi} apply to MaxCut problem as well. Gurobi optimizer is able to generate optimal solutions by modeling the MaxCut problem with integer linear programming \cite{ilp} problem. However, its run time scales exponentially since integer programming is NP-complete. We compare DC-QAOA with classical algorithms on 7 generated graph instances from NetworkX package with varying number of graph size. As depicted in Figure \ref{res}(d), the classical algorithm Random Search shows compromised approximation ratio of 80.73\%, while DC-QAOA achieves 97.14\% which is 20.32\% higher. The optimal MaxCut size calculated from Gurobi optimizer is also shown here as ground truth. The classical algorithm runs significantly faster, within a second for each graph instance, at the expense of highly compromised approximation ratio. Run time of DC-QAOA is displayed in Figure \ref{res}(e).

{\textbf{Quantum counterparts:}} QAOA is unable to efficiently handle large-scale problems. Thus we take quantum annealing method implemented on D-Wave Systems as the quantum counterpart of DC-QAOA. Quantum annealer is the hardware specifically designed for QUBO problems, and its comparison with DC-QAOA implemented on universal quantum computers provides insights on the characteristics of two types of quantum computing techniques. Quantum annealing involves the minor embedding step which maps graph nodes to physical qubits on the hardware. Even though D-Wave 2000Q system supports large qubit size, valid embedding might not be found if an input graph has just few hundreds of nodes but with dense edge connectivity. 
Since the probabilistic nature also exists in quantum annealing, expectation value is compared between QA and DC-QAOA for these graph instances (Figure \ref{res}(e)). Note that QA is able to solve each graph instance within 20 seconds, and run time of quantum annealing largely depends on minor embedding process because the annealing process evolves with quantum advantage. Though DC-QAOA is slower than QA, its run time curve grows with a polynomial rate.

\section{Conclusion}
We propose DC-QAOA to solve large-scale MaxCut 
with exponential speedup compared to conventional QAOA. The proposed Naive \textit{LGP} policy reduces algorithm run time from exponential to polynomial complexity
while the \textit{QSR} policy relies on measurements from sub-solution states and enables recursive state reconstruction for parent quantum states. DC-QAOA generates optimal or near-optimal solutions, depending on the number of measurements on quantum states, while classical algorithms achieve exponential speedup at the expense of compromised approximation ratio. The selection mechanism in \textit{QSR} policy guarantees parent string solutions are reconstructed with high quality, making DC-QAOA superior to quantum annealing in terms of expectation value. 


\section*{Acknowledgement} The work is supported in parts by National Science Foundation (OIA-2040667, CNS-1722557, CCF-1718474, DGE-1723687 and DGE-182176) and seed grants from Penn State Institute for Computational and Data Sciences and Penn State Huck Institute of the Life Sciences.

\bibliographystyle{IEEEtran}
\bibliography{IEEEabrv,ref}

\end{document}